\documentclass[10pt]{article}
\usepackage{fullpage}
\usepackage{amsmath}
\usepackage{amssymb}
\usepackage[dvips]{epsfig}

\advance \textwidth by -2in
\advance \textheight by -3in
\def\##1{{\bf #1}}
\def\=#1{\underline{\underline #1}}
\def\~#1{{\tilde{\bf #1}}}

\def\epso{\epsilon_0}
\def\muo{\mu_0}

\def\.{\mbox{ \tiny{$^\bullet$} }}

\def\curl{\nabla\times}

\def\rt{(\#r,t)}
\def\ro{(\#r,\omega)}

\def\le{\left(}
\def\ri{\right)}
\def\les{\left[}
\def\ris{\right]}

\def\l#1{\label{#1}}

\begin{document}

\noindent Revised for publication in \emph{Chinese Physics Letters}\\
\vspace{5mm} ${}^\dagger$To whom correspondence should be
addressed. Email: T.Mackay@ed.ac.uk

\begin{center}
{\bf {\Large Dyadic Green Function for an Electromagnetic\\  Medium Inspired by
 General
General Relativity }}

 \vspace{5mm}

{ Akhlesh  Lakhtakia}$^1$\footnote{E--mail: akhlesh@psu.edu},
{Tom G. Mackay}$^2$\footnote{Corresponding author. E--mail:
T.Mackay@ed.ac.uk}

\vspace{2mm}

 $^1$\emph{CATMAS, Department of Engineering Science and
Mechanics, Pennsylvania State University, University Park, PA
16802--6812, USA}

 \vspace{2mm}

$^2$\emph{School of Mathematics,
University of Edinburgh, Edinburgh EH9 3JZ, United Kingdom}

\vspace{2mm}

(Received 10 November 2005)

\vspace{2mm}

\end{center}

\noindent \emph{The dyadic Green function for a homogeneous electromagnetic
medium inspired by the spatiotemporally nonhomogeneous constitutive
equations of gravitationally affected vacuum is derived.}

\vspace{2mm}

\noindent \emph{PACS: 41.20.-q, 41.20.Jb, 78.20.-e}

\normalsize \begin{twocolumn}

Vacuum or matter--free space is the most widely studied
electromagnetic medium, not only because it underlies the
development of continuum electromagnetic properties from
microscopic principles,$^{[1,2]}$  but also because of the
significance of electromagnetic communication devices in modern
society.$^{[3]}$ The electromagnetic constitutive equations of
vacuum are commonly stated in textbooks as
\begin{eqnarray}
\label{eq1}
&&\#D(\#r,t) = \epso\, \#E(\#r,t) \,,
\\
\label{eq2}
&&\#B(\#r,t)= \muo\, \#H(\#r,t)\,,
\end{eqnarray}
where $\epso=8.854\times10^{-12}$~F~m$^{-1}$ and
$\muo=4\pi\times10^{-7}$~H~m$^{-1}$ in SI units, whereas $\#r$
and $t$ indicate position and time.

These equations presuppose either the absence of a gravitational
field or that the observer is local. When a gravitational field is
present, spacetime appears curved~---~which is well--
known.$^{[4]}$ One can still use the textbook versions of the
Maxwell postulates for gravitationally affected vacuum, but the
constitutive relations are now$^{[5,6]}$
\begin{eqnarray}
\nonumber
&&\#D\rt = \epso\,\=\gamma\rt\.\#E\rt
\\
\label{eq3}
&&\qquad\quad -\, {c_0^{-1}}\,\#\Gamma\rt\times\#H\rt\,,
\\
\nonumber
&&\#B\rt = \muo\,\=\gamma\rt\.\#H\rt \\
&&\qquad\quad+\, {c_0^{-1}}\,\#\Gamma\rt\times\#E\rt\,,
\label{eq4}
\end{eqnarray}
in lieu of Eqs.\,(1) and (2). Here $\=\gamma\rt$ is a real
symmetric dyadic and $\#\Gamma\rt$ is a vector with real--valued
components, both related to the metric of spacetime; whereas
$c_0=1/\sqrt{\epso\muo}$.

Just as isotropic dielectric--magnetic mediums provide material
counterparts of Eqs.\,(1) and (2) in the frequency
domain,$^{[1,7]}$  the vast variety of complex
materials$^{[8,9]}$~---~natural as well as artificial~---~suggests
that it is quite possible that  Eqs.\,(3) and (4) also have
material counterparts. This thought inspired the present Letter,
wherein we present the derivation of the dyadic Green function for
frequency--domain electromagnetic fields in a homogeneous
\linebreak medium inspired by Eqs.\,(3) and (4).

With the assumption that all fields have an $\exp(-i\omega t)$ time
dependence, with $\omega$ as the angular frequency,
the constitutive relations of the chosen medium are
\begin{eqnarray}
\nonumber
&&\#D\ro = \epso\,\=\gamma(\omega)\.\#E\ro\\
&& \qquad\quad-\, {c_0^{-1}}\,\#\Gamma(\omega)\times\#H\ro\,,
\label{eq5}
\\
\nonumber
&&\#B\ro = \muo\,\=\gamma(\omega)\.\#H\ro
\\
&&\qquad\quad+ \,c^{-1}_0 \,\#\Gamma(\omega)\times\#E\ro\,.
\label{eq6}
\end{eqnarray}
 The coordinate system has been chosen such that
$\=\gamma(\omega)$ is diagonal, and from now onwards the
dependence on $\omega$ is implicit. Let us stress that  Eqs.\,(5)
and  (6) are taken here to describe a material medium which can be
potentially be fabricated in a laboratory by properly dispersing
electrically small bent--wire and other complex inclusions of
different shapes and materials in some host material,$^{[10-13]}$
but should not be confused with the constitutive equations (3) and
(4) of gravitationally  affected vacuum.

The frequency--domain Maxwell curl postulates in the chosen medium may be
set down
down as
\begin{eqnarray}
\nonumber
&&\curl\#E(\#r)=i\omega\les \muo\,\=\gamma\.\#H(\#r) + {c_0^{-1}}\,\#\Gamma\times\#E(\#r)\ris \,,\\
&&
\label{Eeqn}
\\
\nonumber
&&\curl\#H(\#r)=-i\omega\les
\epso\,\=\gamma\.\#E(\#r) - {c_0^{-1}}\,\#\Gamma\times\#H(\#r)\ris
\\
&&\hspace{20mm}
+\,\#J(\#r)\,,
\label{Heqn}
\end{eqnarray}
where $\#J(\#r)$ is the source electric current density.
Our objective is to find the dyadic Green functions
$\=G_{\,e}(\#r,\#s)$ and $\=G_{\,m}(\#r,\#s)$ such that
\begin{eqnarray}
\label{Esol} \#E(\#r) &=&
i\omega\muo\int\int\int\,\=G_{\,e}(\#r,\#s)\.\#J(\#s)\,d^3\#s\,,
\\
\label{Hsol}
 \#H(\#r)
&=&\int\int\int\,\=G_{\,m}(\#r,\#s)\.\#J(\#s)\,d^3\#s\,,
\end{eqnarray}
with  the integrations being  carried out over the region where
the source electric current density is nonzero.

To begin with, the substitution of Eq.\,(9) into Eq.\,(7) and
comparison of the resulting expression with Eq.\,(10)  yields
\begin{equation}
\=G_{\,m}(\#r,\#s) = \=\gamma^{-1}\.\left(\curl\=I
-ik_0\,\#\Gamma\times\=I\right)\. \=G_{\,e}(\#r,\#s)\,,
\end{equation}
where $k_0=\omega\sqrt{\epso\muo}$ and $\=I$ is the identity
dyadic. Thus, an expression for only $\=G_{\,e}(\#r,\#s)$ has to
be found.

For that purpose, following Lakhtakia and \linebreak
Weiglhofer,$^{[14]}$ we start by defining new fields and source
current density as
\begin{eqnarray}
&&\#e(\#r)=\#E(\#r)\,\exp(-ik_0\,\#\Gamma\.\#r)\,, \l{e_LW}
\\
&&\#h(\#r)=\#H(\#r)\,\exp(-ik_0\,\#\Gamma\.\#r)\,,
\\
&&\#j(\#r)=\#J(\#r)\,\exp(-ik_0\,\#\Gamma\.\#r)\,. \l{j_LW}
\end{eqnarray}

Hence, Eqs.\,(7) and (8) respectively transform to
\begin{eqnarray}
\label{eeqn}
&&\curl\#e(\#r)=i\omega  \muo\,\=\gamma\.\#h(\#r)  \,,
\\
\label{heqn}
&&\curl\#h(\#r)=-i\omega
\epso\,\=\gamma\.\#e(\#r)
+\#j(\#r)\,.
\end{eqnarray}

Next, we make use of an affine transformation associated with the
scaling of space as per$^{[15]}$
\begin{equation}
\~r=\=\gamma^{1/2}\.\#r\,,
\end{equation}
where $ \=\gamma^{1/2} \. \=\gamma^{1/2} = \=\gamma  $  and
we recall that $\=\gamma $
is a dyadic with real--valued
elements. Let us define
another set of fields and source current density as
\begin{eqnarray}
&&\~e(\#r) = \=\gamma^{1/2}\.\#e(\=\gamma^{1/2}\.\#r)\,,
\l{e_affine}
\\
&&\~h(\#r) = \=\gamma^{1/2}\.\#h(\=\gamma^{1/2}\.\#r)\,,
\\
&&\~j(\#r) = \left({\rm
adj}\,\=\gamma^{1/2}\right)\.\#j(\=\gamma^{1/2}\.\#r)\,,
\l{j_affine}
\end{eqnarray}
where `adj'  stands for the adjoint. Then, Eqs.\,(15) and (16)
transform to
\begin{eqnarray}
\label{eeeqn}
&&\curl\~e(\#r)=i\omega  \muo\, g\,\~h(\#r)  \,,
\\
\label{hheqn} &&\curl\~h(\#r)=-i\omega \epso\,g\,\~e(\#r)
+\~j(\#r)\,,
\end{eqnarray}
where
\begin{equation}
g=\sqrt{\vert\=\gamma\vert}\,,
\end{equation}
and $\vert\=\gamma\vert$ denotes the determinant of $\=\gamma $.

From the foregoing equations, we obtain
\begin{eqnarray}
\nonumber
&&
\les(\curl\=I)\.(\curl\=I)-k_0^2g^2\=I\ris\.\~e(\#r)\\
&&\qquad\quad =i\omega\muo\,g\,\~j(\#r)\,.
\label{final-e}
\end{eqnarray}
The solution of Eq.\,(24) is well--known as$^{[16]}$
\begin{equation}
\label{eesol}
\~e(\#r)=i\omega\muo\,g\,
\int\int\int \tilde{\=g}(\#r,\#s)\.\~j(\#s)\,d^3\#s\,,
\end{equation}
where
\begin{equation}
\tilde{\=g}(\#r,\#s)= \left(\=I
+\frac{\nabla\nabla}{k_0^2g^2}\right)\,
\frac{\exp(ik_0g\vert\#r-\#s\vert )}{4\pi\vert\#r-\#s\vert}\,.
\end{equation}

In order to go back from Eq.\,(25)  to  Eq.\,(9), we have to
invert the two transformations in reverse sequence: Substitution
of Eqs.\,(18) and (20) in Eq.\,(25) yields
\begin{eqnarray}
\nonumber
&&
\#e(\#r)=i\omega\muo\, \left({\rm adj}\,\=\gamma^{1/2}
\, \right)  \. \left( \int\int\int {{\=g}}(\#r, \#s)\right.
\\
&&\qquad 
\.\#j(\#s)\,d^3\#s\,
\Big) \. \left({\rm adj}\,\=\gamma^{1/2} \,  \right) ,  \l{rev_1}
\end{eqnarray}
wherein
\begin{eqnarray}
\nonumber
&&
 {{\=g}}(\#r,\#s)= \left(\=I +\frac{1}{k_0^2g^2} \=\gamma^{1/2} \.
\nabla\nabla  \. \=\gamma^{1/2} \right) \\
&&\quad\frac{\exp\les ik_0g\vert
\=\gamma^{-1/2} \. \le  \#r-\#s \ri \vert \ris}{4\pi\vert
\=\gamma^{-1/2} \. \le \#r-\#s \ri \vert}\,,\end{eqnarray}
By substituting for $\#{\it\bf e}$ and $\#{\it\bf j}$ in Eq.\,(27)
using Eqs.\,(12) and (14), respectively, we find
\begin{eqnarray}
\nonumber
&&
\#E(\#r) =  i\omega\muo\,\exp (i k_0 \#\Gamma \. \#r
)\,\left({\rm adj}\,\=\gamma^{1/2} \, \right)  \\
\nonumber
&&\quad\. \Big( \int\int\int
{{\=g}}(\#r, \#s)\.\#J(\#s) \exp ( - i k_0 \#\Gamma \. \#s )
\,d^3\#s\, \Big)\\
&&\qquad \. \left({\rm adj}\,\=\gamma^{1/2} \, \right)\,.
 \l{rev_2}
\end{eqnarray}
Therefore, the dyadic Green function $\=G_{\,e}(\#r,\#s)$ emerges
from  Eq.\,(29) as
\begin{eqnarray}
\nonumber
 &&\=G_{\,e}(\#r,\#s) = \exp \les i k_0 \#\Gamma \. \le \#r - \#s
\ri \ris\, \left( {\rm adj}\,\=\gamma  +\frac{1}{k_0^2}
 \nabla\nabla  \right) \\
 &&\qquad \frac{\exp\les ik_0g\vert \=\gamma^{-1/2} \. \le \#r-\#s
\ri \vert \ris}{4\pi\vert \=\gamma^{-1/2} \. \le \#r-\#s \ri \vert}
\, . \label{eq30}
\end{eqnarray}

Equation (30) is the desired result. If $\=\gamma=\=I$ and
$\#\Gamma=\#0$, this expression reduces to the usual dyadic Green
function for gravitationally unaffected vacuum.$^{[16]}$

\end{twocolumn}
\end{document}